\documentstyle[twocolumn,pra,aps]{revtex}
\def\begmc{}
\def\endmc{}

\def\Im{\mathop{\rm Im}}
\def\Re{\mathop{\rm Re}}

\begin{document}
\title{Causality, stability and passivity for a mirror in vacuum}
\author{Marc Thierry Jaekel $^{(a)}$ and Serge Reynaud $^{(b)}$}
\address{(a) Laboratoire de Physique Th\'{e}orique de l'ENS
\thanks{%
Unit\'e propre du Centre National de la Recherche Scientifique,
associ\'ee \`a l'Ecole Normale Sup\'erieure et \`a l'Universit\'e
Paris-Sud}, 24 rue Lhomond, F75231 Paris Cedex 05 France\\
(b) Laboratoire de Spectroscopie Hertzienne
\thanks{%
Unit\'e de l'Ecole Normale Sup\'erieure et de l'Universit\'e
Pierre et Marie Curie, associ\'ee au Centre National de la Recherche
Scientifique}, 4 place Jussieu, case 74, F75252 Paris Cedex 05 France}
\date{{\sc Physics Letters} {\bf A 167} (1992) 227-232}
\maketitle

\begin{abstract}
The mean force exerted upon a perfect mirror moving in vacuum in a two
dimensional spacetime has the same expression as the radiation reaction
force computed in classical electron theory. It follows that unacceptable
runaway solutions are predicted. We show that this instability problem does
not appear when partially transmitting mirrors are studied. The mechanical
impedance describing the mirror coupled to vacuum radiation pressure is
computed explicitly; recoil is neglected. It is found to be a passive
function, so that stability is ensured. This is connected to the fact that
no energy can be extracted from the vacuum state.
\end{abstract}

\begmc

\section*{Introduction}

A mirror moving in vacuum experiences a mean force which is associated
through linear response theory with the quantum fluctuations of the vacuum
stress tensor \cite{Passiv1}.

This motional force, denoted hereafter $F_{m}$, has been computed by Fulling
and Davies \cite{Passiv2} for a perfectly reflecting mirror scattering a
scalar vacuum field in a two dimensional spacetime. A linear approximation
in the mirror's displacement leads from their result to a force proportional
to the third derivative of the time dependent mirror's position $q$ 
\begin{equation}
F_{m}(t)=\frac{\hbar q^{\prime \prime \prime }(t)}{6\pi c^2 }  \eqnum1 
\end{equation}
As required by Lorentz invariance of the vacuum state, this force vanishes
for a uniform velocity. It also vanishes for a uniform acceleration so that
the inertial mass of the mirror is not modified.

The motional force modifies the mechanical response of the mirror to an
applied force $F_{a}$ since the equation of motion can be written 
\begin{equation}
kq(t)+mq^{\prime \prime }(t)=F_{a}(t)+F_{m}(t)\qquad k=m\omega _0 ^2  
\eqnum2 
\end{equation}
where $m$ is the mass of the mirror, $\omega _0 $ is the eigenfrequency of
the suspension system ($\omega _0 =0$ in the limiting case of a free mass)
and $F_{m}$ is the motional force. Consequently, it has to be taken into
account in a complete analysis of the sensitivity of interferometric
measurements. This seems particularly important for interferometric
detection of gravitational waves \cite{Passiv3}.

The force (1) has the same expression, with a modification of the numerical
factor, as the radiation reaction force computed in classical electron
theory \cite{Passiv4}. It is well known that severe problems are associated
with the resulting equation of motion (2): a historical account is given in
ref. \cite{Passiv4}; more recent references can be found in \cite{Passiv5}.
First, it turns out that `runaway solutions' exist, which correspond to an
exponential increase of the velocity in the absence of applied force. These
runaway solutions may be prohibited by a special choice of boundary
conditions, but a `preacceleration' phenomenon then occurs, with an
acceleration of the mass before the force is applied \cite{Passiv4}.

Clearly, the equation of motion (2) violates causality for a perfect mirror
when the motional force is given by equation (1). We show in this letter
that this problem is solved in a simple manner by considering partially
transmitting mirrors rather than perfect ones. We will assume that the
mirror is transparent at frequencies higher than a reflection cutoff $\omega
_{C}$ and that this cutoff obeys the following inequality 
\begin{equation}
\hbar \omega _{C}\ll mc^2   \eqnum3 
\end{equation}
This condition, obeyed by realistic mirrors ($m$ is a macroscopic mass),
implies that the recoil effect for reflected fields can be ignored in the
derivation of the motional force \footnote{%
The perfect mirror corresponds to $\omega _{C}\rightarrow \infty $ and does
not obey condition (3). It is therefore necessary to take the recoil into
account, which was not done in the derivation of the force (1). This
probably explains the inconsistencies associated with expression (1) of the
force.}.

Using a previously derived expression for the motional force associated with
a partially transmitting mirror \cite{Passiv1}, we will demonstrate that the
mirror coupled to the vacuum radiation pressure is a stable system (runaway
solutions disappear). More precisely, we will show that the mirror obeys
`passivity' \cite{Passiv6} when some inequality which is a consequence of
condition (3) is satisfied. Passivity is closely connected to energy
considerations: the moving mirror radiates energy into vacuum and is
therefore damped, while the vacuum cannot provide energy for exciting the
mirror's motion. Then, stability follows and problems associated with
runaway solutions are eliminated.

\section*{The susceptibility function associated with the motional force}

At the limit of small displacements, the motional force is a function of $q$
easily defined in the frequency domain in terms of a linear susceptibility 
$\chi $ 
\begin{eqnarray}
&&F_{m}(t)={\int }{\rm d}t^\prime \chi (t-t^\prime )q(t^\prime ) 
\eqnum{4a} \\
&&F_{m}[\omega ]=\chi [\omega ]q[\omega ]  \eqnum{4b}
\end{eqnarray}
Any function $f$ is written in the time domain or in the frequency domain
according to the rule 
\[
f(t)={\int }\frac{{\rm d}\omega }{2\pi }f[\omega ]e^{-i\omega t}
\]

The motional force (1) computed for a perfect mirror corresponds to the
susceptibility 
\begin{equation}
\chi [\omega ]=\frac{i\hbar \omega ^3 }{6\pi c^2 }=im\tau \omega
^3 \qquad \tau =\frac{\hbar }{6\pi mc^2 }  \eqnum{5}
\end{equation}
where $\tau $ is a time constant, characteristic of the coupling of the
mirror of mass $m$ to the vacuum radiation pressure. The corresponding time
constant for the electron is of the order of 10$^{-22}$s. It is much smaller
for a macroscopic mirror.

In a scattering approach \cite{Passiv7}, a partially transmitting mirror is
described by a reflectivity function $r[\omega ]$ and a transmitivity
function $s[\omega ]$, which obey unitarity and causality conditions. The
mirror is supposed to become transparent at the high frequency limit 
($r\rightarrow 0$ when $\omega \rightarrow \infty $). The perfect mirror is
then replaced by the more realistic model of a causal mirror having a
perfect reflection below a reflection cutoff $\omega _{C}$, but transparent
at higher frequencies.

This approach has provided us with an expression for the mean Casimir force
between two motionless mirrors \cite{Passiv7}, which is free from the
divergences usually associated with the infiniteness of the vacuum energy,
as well as a causal expression for the motional Casimir force between moving
mirrors \cite{Passiv8}.

When condition (3) is obeyed, it is possible to ignore the recoil effect and
the motional force can be written as \footnote{%
The motional force for a mirror in vacuum is given by equations (33) and
(11) of ref. \cite{Passiv1}} 
\begin{equation}
\chi [\omega ]=im\tau \omega ^3 \Gamma [\omega ]  \eqnum{6a}
\end{equation}
where 
\begin{eqnarray}
&&\omega ^3 \Gamma [\omega ] = {\int_0 ^\omega }3{\rm d}\omega ^\prime
\ (\omega -\omega ^\prime )\omega ^\prime \alpha [\omega -\omega
^\prime ,\omega ^\prime ]  \eqnum{6b} \\
&&\alpha [\omega ,\omega ^\prime ] = 1-s[\omega ]s[\omega ^\prime
]+r[\omega ]r[\omega ^\prime ]  \eqnum{6c}
\end{eqnarray}
The function $\Gamma [\omega ]$ is regular at the low frequency limit 
\[
\Gamma [\omega ]\rightarrow \Gamma [0]=r[0]^2 \qquad {\rm for\ }\omega
\rightarrow 0
\]
When $r[0]=-1$, the susceptibility (5) is recovered at low frequencies
whatever the reflectivity may be at high frequencies. At high frequencies, 
$\Gamma [\omega ]$ behaves as a cutoff function. Its behaviour will be
discussed in more detail later on.

A simple example is provided by the following model fulfilling unitarity,
causality and transparency conditions 
\begin{equation}
r[\omega ]=\frac{-1}{1-\frac{i\omega }\Omega }\qquad s[\omega ]=1+r[\omega
]  \eqnum{7a}
\end{equation}
One obtains in this case 
\begin{eqnarray}
\chi [\omega ] &=&6m\tau \Omega ^3   \nonumber \\
&\times &\left( -\frac{i\omega }\Omega -\frac{\omega ^2 }{2\Omega ^2 }
-\left( 1-\frac{i\omega }\Omega \right) \ln \left( 1-\frac{i\omega }
\Omega \right) \right)   \nonumber \\
\Gamma [\omega ] &=&6\left( \frac 1 {2.3}+\frac 1 {3.4}\frac{i\omega }
\Omega +\frac 1 {4.5}\left( \frac{i\omega }\Omega \right) ^2 +\ldots
\right)   \nonumber \\
&&\qquad {\rm for\ }\omega \ll \Omega   \eqnum{7b}
\end{eqnarray}

\section*{Positivity of the dissipative functions}

Coming back to the general case, we introduce the real and imaginary parts
of the functions written in the frequency domain, which are associated with
dissipation and dispersion 
\begin{eqnarray*}
&&\chi [\omega ] = \chi _{R}[\omega ]+i\chi _{I}[\omega ]\qquad \Gamma
[\omega ]=\Gamma _{R}[\omega ]+i\Gamma _{I}[\omega ] \\
&&\chi _{I}[\omega ] = m\tau \omega ^3 \Gamma _{R}[\omega ]\qquad \chi
_{R}[\omega ]=-m\tau \omega ^3 \Gamma _{I}[\omega ]
\end{eqnarray*}
$\chi _{R}$ and $\Gamma _{R}$ are even functions of $\omega $ while $\chi
_{I}$ and $\Gamma _{I}$ are odd functions of $\omega $ (see eqs 4, 6). The
dissipative part $\chi _{I}$ of the susceptibility, which was denoted $\xi
_{FF}$ in ref. \cite{Passiv1}, is the commutator of the force operator.

Using the unitarity of the $S-$matrix associated with the mirror \cite
{Passiv1}, one shows that 
\begin{eqnarray*}
2\Re \alpha [\omega ,\omega ^\prime ] &=&|\alpha [\omega ,\omega
^\prime ]|^2 +|\beta [\omega ,\omega ^\prime ]|^2  \\
\beta [\omega ,\omega ^\prime ] &=&s[\omega ]r[\omega ^\prime ]-r[\omega
]s[\omega ^\prime ]
\end{eqnarray*}
The following positivity property follows 
\begin{eqnarray}
\Gamma _{R}[\omega ] &\geq &0\qquad {\rm for\ }\omega \ \rm real
\nonumber \\
\chi _{I}[\omega ] &\geq &0\qquad {\rm for\ }\omega \ \rm real\ 
\geq 0  \eqnum{8}
\end{eqnarray}
This property corresponds to the dissipative character (see for example ref. 
\cite{Passiv9}) of the motional perturbation in the vacuum and will play an
important role in the following.

\section*{Dispersion relations}

Causality \cite{Passiv9} is associated with the fact that the susceptibility
function, written in the time domain (see eq 4), vanishes for negative times 
\[
\chi (t)=0\qquad {\rm for\ }t<0
\]
In the frequency domain, a causal function is analytic and regular in the
half plane $\Im \omega >0$. For the susceptibility given by equations
(6), these properties follow \cite{Passiv1} from the causality of the
scattering coefficients $r$ and $s$.

The real and imaginary parts of a causal function are related through
dispersion relations, which take the simple form of Kramers-Kronig relations
when the function decreases sufficiently rapidly at high frequencies \cite
{Passiv9}. But this is not the case for the susceptibility function $\chi $,
so that it is necessary to write dispersion relations with subtractions \cite
{Passiv10}.

From equations (6), one knows that the three quantities $\chi [0]$, $\chi
^\prime [0]$ and $\chi ^{\prime \prime }[0]$ vanish. It is therefore
tempting to write the dispersion relation with three subtractions at $\omega
=0$ which is just the Kramers-Kronig relation for the function $\Gamma $ 
\begin{equation}
\Gamma [\omega ]={\int }\frac{{\rm d}\omega ^\prime }{i\pi }\frac{\Gamma
_{R}[\omega ^\prime ]}{\omega ^\prime -\omega -i\epsilon }  \eqnum{9}
\end{equation}
Assuming that $\Gamma [\omega ]$ decreases sufficiently rapidly at high
frequencies so that it is a square integrable function, the relation (9) is
actually equivalent to causality for the susceptibility function. This
follows from the Titchmarsh theorem \cite{Passiv10}.

As $\Gamma _{R}$ is positive, one deduces that $\Gamma [\omega ]$ behaves as 
$\frac 1 \omega $ at high frequencies 
\begin{equation}
\Gamma [\omega ]\approx \frac{\omega _{C}}{-i\omega }\qquad {\rm for\ }
\omega \rightarrow \infty   \eqnum{10}
\end{equation}
where the parameter $\omega _{C}$ can be considered as the definition of the
reflection cutoff (we suppose that the integral is finite) 
\begin{equation}
\omega _{C}={\int }\frac{{\rm d}\omega }{\pi }\Gamma _{R}[\omega ]=\int 
\frac{{\rm d}\omega }{\pi }\frac{\chi _{I}[\omega ]}{m\tau \omega ^3 } 
\eqnum{11}
\end{equation}
For example, the model of equations (7) corresponds to 
$\omega _{C}=3\Omega $.

The behaviour of the susceptibility is described by a high frequency mass 
$\mu $ 
\begin{eqnarray}
\chi [\omega ] &\approx &-\mu \omega ^2 \qquad {\rm for\ }\omega
\rightarrow \infty   \nonumber \\
\mu  &=&m\omega _{C}\tau =\frac{\hbar \omega _{C}}{6\pi c^2 }  \eqnum{12}
\end{eqnarray}
The induced mass $\mu $ is quite similar to the `electromagnetic mass' of
electron theory \cite{Passiv4}. It is positive (see eq 8) and much smaller
than $m$ for realistic macroscopic mirrors (see eq 3).

The dispersion relation (9) can be written equivalently in the time domain 
\[
\Gamma (t)=2\theta (t)\Gamma _{R}(t)
\]
It is thus clear that the susceptibility function does not obey the same
simple dispersion relation; one can indeed write 
\begin{eqnarray*}
\chi (t) =2m \tau &&\left( \theta (t)\Gamma _{R}^{\prime \prime \prime
}(t)+\delta (t)\Gamma _{R}^{\prime \prime }(0)\right.  \\
&&+\left. \delta ^\prime (t)\Gamma _{R}^\prime (0)+\delta ^{\prime
\prime }(t)\Gamma _{R}(0)\right) 
\end{eqnarray*}
It follows that the expression (4a) of the motional force cannot be written
simply in terms of the force commutator $\chi _{I}(t)$ since quasistatic
motions must be subtracted up to the second order in a Mac-Laurin expansion.
However, using the parity property of the function $\Gamma _{R}$, one
recovers the consistency condition of linear response theory \footnote{%
The dispersion relation written in refs. \cite{Passiv1} and \cite{Passiv8}
did not account for subtractions and has therefore to be amended. However,
this does not affect the expressions for the linear susceptibility, the
force commutator and the force fluctuations which were independently
computed. Also, the conclusion that these quantities are consistent with
linear response theory remains valid.} 
\[
\chi (t)-\chi (-t)=2m\tau \Gamma _{R}^{\prime \prime \prime }(t)=2i\chi
_{I}(t)
\]

\section*{Mechanical impedance of the mirror}

When the coupling of the mirror to vacuum radiation pressure is taken into
account, the equation of motion (2) is easily solved in the frequency domain.

The mechanical impedance $Z$ gives the frequency components of the force 
$F_{a}$ as a function of the velocity $v$ 
\begin{eqnarray}
v(t) &=&q^\prime (t)\qquad v[\omega ]=-i\omega q[\omega ]  \nonumber \\
F_{a}[\omega ] &=&Z[\omega ]v[\omega ]  \nonumber \\
-i\omega Z[\omega ] &=&k-m\omega ^2 -\chi [\omega ]  \eqnum{13}
\end{eqnarray}
The two real functions $Z_{R}$ and $Z_{I}$ are respectively the dissipative
and dispersive parts of the impedance 
\begin{eqnarray}
Z[\omega ] &=&Z_{R}[\omega ]+iZ_{I}[\omega ]  \nonumber \\
Z_{R}[\omega ] &=&m\tau \omega ^2 \Gamma _{R}[\omega ]  \nonumber \\
Z_{I}[\omega ] &=&\frac{k}\omega -m\omega +m\tau \omega ^2 \Gamma
_{I}[\omega ]  \eqnum{14}
\end{eqnarray}

Equivalently, we can write the velocity $v$ as a function of the applied
force $F_{a}$ and of the mechanical admittance $Y$ 
\[
v[\omega ]=Y[\omega ]F_{a}[\omega ]\qquad Y[\omega ]=\frac 1 {Z[\omega ]}
\]

\section*{Stability}

From causality, we know that the motional force is a retarded function of
the mirror's position, which implies that the impedance function $Z[\omega ]$
is analytic and has no poles in the half plane $\Im \omega >0$.

The analytic properties of the admittance function $Y[\omega ]$ require a
closer examination since its expression in terms of the susceptibility $\chi
[\omega ]$ has the form of a `closed loop gain', so that a self oscillation
could occur. Such a self oscillation would be associated with the presence
of a pole for $Y$, that is a zero for $Z$, in the half plane $\Im \omega
>0$.

But, if the suspended mirror coupled to the vacuum is a stable system, the
self oscillation cannot take place. It follows that the admittance function 
$Y[\omega ]$ shall also be analytic and regular in the half plane $\Im 
\omega >0$ and that the impedance function $Z[\omega ]$ should have no zero
in this domain.

Using the susceptibility function (5) of a perfect mirror (with neglected
recoil) and assuming that the mirror is free ($k=0$), one checks that the
function $Y$ has a pole in the half plane $\Im \omega >0$, at $\omega =
\frac{i}{\tau }$, in this case. This is similar to the well known
instability problem of electron theory \cite{Passiv4}.

This result is modified when a partially transmitting mirror is studied.
Indeed, the pole associated with runaway solutions corresponds to a
frequency well above the reflection cutoff (see eq 3). Due to the asymptotic
behaviour of the cutoff function $\Gamma $ (see eq 10), the pole which was
located at $\omega =\frac{i}{\tau }$ for a perfect mirror is shifted to 
$\omega \approx -i\omega _{C}$ so that the instability is suppressed.

\section*{Passivity}

We demonstrate now in a more rigorous manner that the instability is
suppressed for any reflectivity function, provided that the following
inequality is obeyed 
\begin{equation}
\mu \leq m  \eqnum{15}
\end{equation}
In particular, this will be the case as soon as the cutoff fulfills
condition (3).

The demonstration relies upon the dispersion relation (9) which reads for
the impedance function, for $\Im \omega >0$ 
\[
Z[\omega ]=\omega ^2 {\int }\frac{{\rm d}\omega ^\prime }{\pi }\frac{
Z_{R}[\omega ^\prime ]}{\omega ^{\prime \ 2}\left( \epsilon -i\omega
+i\omega ^\prime \right) }+\frac{ik}\omega -im\omega 
\]
Using Laplace transforms instead of Fourier transforms 
\[
f\{p\}=f[ip]\qquad \Re p>0
\]
this relation can be transformed to 
\begin{equation}
Z\{p\}={\int }{\rm d}\Phi (\rho )\frac{1-ip\rho }{p-i\rho }+\frac{k}{p}
+p(m-\mu )  \eqnum{16}
\end{equation}
with $\mu $ defined by equations (12) and (11) and the spectral function 
$\Phi $ given by 
\[
{\rm d}\Phi (\rho )=\frac{Z_{R}[\rho ]{\rm d}\rho }{\pi \left( 1+\rho
^2 \right) }
\]
One knows that $Z_{R}$ is positive at all frequencies (see eqs 8 and 14),
that the function $\Phi $ is bounded (this follows from the finiteness of 
$\mu $) and that the coefficients $k$ and $m-\mu $ are positive (when $\mu
\leq m$). Then, equation (16) is actually the spectral representation for a
passive function, that is a function obeying the property \cite{Passiv6} 
\begin{equation}
\Re Z\{p\}\geq 0\qquad {\rm for} \ \Re p>0  \eqnum{17}
\end{equation}
Causality and stability follow from passivity \cite{Passiv10}, so that the
problems associated with runaway solutions are eliminated.

It has to be emphasized that the passivity condition (17) is more stringent
than the already known positivity property (8): as an example, the
expressions corresponding to the perfect mirror obey condition (8) but not
condition (17).

It is also worth to stress that the passivity property is obtained only when
the mass $m$ of the mirror is greater than the induced mass $\mu $ which is
similar to an electromagnetic mass. As far as electron theory is concerned,
the same conclusion is reached by Dekker \cite{Passiv11}. In other words,
passivity is obeyed by the impedance $Z$ but not by the contribution of the
motional susceptibility taken separately \footnote{%
At low frequencies $\frac{\chi [\omega ]}{i\omega }\approx m\tau \omega
^2 \Gamma [0]$, so that this function does not obey the passivity property
(17) in the vicinity of $\omega =0$.}, in contrast with the usual models of
Brownian motion where the motional contribution itself is a passive function 
\cite{Passiv6,Passiv12}.

\section*{Energy considerations}

Passivity is equivalent to the proposition that a positive amount of energy
is transfered from the reservoir (source of the applied force $F_{a}$) to
the moving mirror \cite{Passiv6} 
\begin{equation}
W_{a}(t)={\int_{-\infty }^{t}}{\rm d}t^\prime \ F_{a}(t^\prime)
v(t^\prime )\geq 0  \eqnum{18}
\end{equation}

One immediately deduces from the equation of motion (2) that 
\begin{eqnarray*}
W_{a}(t) &=&\Delta E(t)+W_{m}(t) \\
\Delta E(t) &=&E(t)-E(-\infty ) \\
E(t) &=&\frac 1 2 kq(t)^2 +\frac 1 2 mv(t)^2  \\
W_{m}(t) &=&-{\int_{-\infty }^{t}}{\rm d}t^\prime \ F_{m}(t^\prime)
v(t^\prime )
\end{eqnarray*}
The energy $W_{a}$ provided by the reservoir is the sum of an energy $\Delta
E$ stored in the mirror's motion and of a part $W_{m}$ radiated in the
vacuum state.

When we consider a situation where the force $F_{a}$ is applied only during
a limited time interval \cite{Passiv2} and where the initial and final
states of the mirror are the same ($\Delta E(\infty )=0$), we conclude from
condition (18) that the net radiated energy is positive 
\[
W_{a}(\infty )=W_{m}(\infty )\geq 0
\]
Thus, passivity appears to be connected to the fact that vacuum can damp the
mirror's motion but cannot excite it. This is the ultimate reason why
stability follows from passivity: the vacuum cannot provide energy for
sustaining runaway solutions.

\section*{Conclusion}

The expression (1) of the damping force for a perfect mirror in vacuum is
associated with unacceptable runaway solutions. We have shown that this
problem does not take place for a realistic mirror with a reflectivity
cutoff obeying condition (3).

The impedance function can be computed explicitly, neglecting the recoil
effect, and is found to be a passive function, so that stability is ensured.
This is connected to the fact that no energy can be extracted from the
vacuum state.

It is tempting to consider that passivity is actually a general property of
the impedance function for a scatterer in vacuum. Then, a consistent
treatment of a perfect mirror, taking into account the recoil effect for
reflected fields, would necessarily provide us with a passive function. This
would also be true for a satisfactory treatment of the electron. Attempts in
this direction, unsuccessful up to now, are discussed by Rueda 
\cite{Passiv5}.

\endmc

\end{document}